\documentclass[runningheads]{llncs}
\usepackage{subfig}
\usepackage{graphicx}
\usepackage{multirow}
\usepackage{todonotes}
\usepackage{booktabs}
\usepackage{tabularx}
\usepackage{paralist}

\usepackage[misc,geometry]{ifsym} 

\newcolumntype{R}{>{\raggedleft\arraybackslash}X}

\begin{document}
\title{How do Quantifiers Affect the Quality of Requirements?}
%
%
\author{Katharina Winter\textsuperscript{(\Letter)}  \inst{1}
\and Henning Femmer \orcidID{0000-0002-6059-4635} \inst{2}
\and Andreas Vogelsang\textsuperscript{(\Letter)} \orcidID{0000-0003-1041-0815} \inst{3}
}
\authorrunning{K. Winter et al.}
%

\institute{
Technische Universit{\"a}t M{\"u}nchen, Germany\\
\email{kathi.winter@tum.de},\\ 
\and
Qualicen GmbH \\
Garching, Germany\\
\email{henning.femmer@qualicen.de}\\
\and
Technische Universit{\"a}t Berlin, Germany\\
\email{andreas.vogelsang@tu-berlin.de}\\ 
}

\maketitle              
\begin{abstract}
\textbf{[Context]} Requirements quality can have a substantial impact on the effectiveness and efficiency of using requirements artifacts in a development process. Quantifiers such as ``at least'', ``all'', or ``exactly'' are common language constructs used to express requirements. Quantifiers can be formulated by affirmative phrases (``At least'') or negative phrases (``Not less than'').
\textbf{[Problem]} It is long assumed that negation in quantification negatively affects the readability of requirements, however, empirical research on these topics remains sparse.
\textbf{[Principal Idea]} In a web-based experiment with 51 participants, we compare the impact of negations and quantifiers on readability in terms of reading effort, reading error rate and perceived reading difficulty of requirements.
\textbf{[Results]} For 5 out of 9 quantifiers, our participants performed better on the affirmative phrase compared to the negative phrase. Only for one quantifier, the negative phrase was more effective.
\textbf{[Contribution]} This research focuses on creating an empirical understanding of the effect of language in Requirements Engineering. It furthermore provides concrete advice on how to phrase requirements.

\end{abstract}

\section{Introduction}
\label{chp:Intro}
Requirements are a crucial part of the software development process. However, in contrast to the code making up the software, requirements themselves do not have much direct value for a customer. Femmer and Vogelsang define requirements as ``means for a software engineering project''~\cite{Femmer19}. Thus, bad quality in requirements may result in issues that possibly arise in later stages of the development process leading to a rework of process steps, potentially impacting software code or tests, for example. Indicators of these potential quality issues are named ``Requirements Smells''~\cite{Femmer17}, including, for instance, ambiguous words or passive voice. In this paper, we examine the use of specific quantifiers as one particular type of requirements smells. Although the use of quantifiers, such as ``at least'', ``all'', or ``exactly'', is substantial in requirements specifications~\cite{Berry05}, they have not received much attention in literature so far. Questions on how different use and phrasing of quantifiers affect the quality of requirement artifacts remain unacknowledged. To shed light on this topic, we categorize the quantifiers into different scopes and use this categorization as a theoretical foundation to compare them. Each quantifier scope has one semantic interpretation but can be expressed in different syntactic ways. For example, ``At least'' and ``Not less than'', belong to the same semantic scope but one is expressed in an affirmative syntax, while the other is expressed in negative syntax.

In this paper, we examine 9 different quantifier scopes and compare the impact on requirements readability. We conducted an experiment with 51 participants and compare reading times, error rates, and perceived difficulty of quantifiers in affirmative and negative syntax. The goal of our research is to provide empirical evidence for justifying requirements writing guidelines and offer best practices on quantifier usage in requirements specifications. 

Our results show that the use and phrasing of specific quantifiers has a significant effect on reading times, errors, and perceived difficulty. Based on our results, we formulate concrete advice for writing better requirements.

\section{Background}

\subsection{Quantifiers in the English Language}
\label{sec:1:Quantifiers}
\begin{figure}
	\centering
	\includegraphics[width=0.70\textwidth]{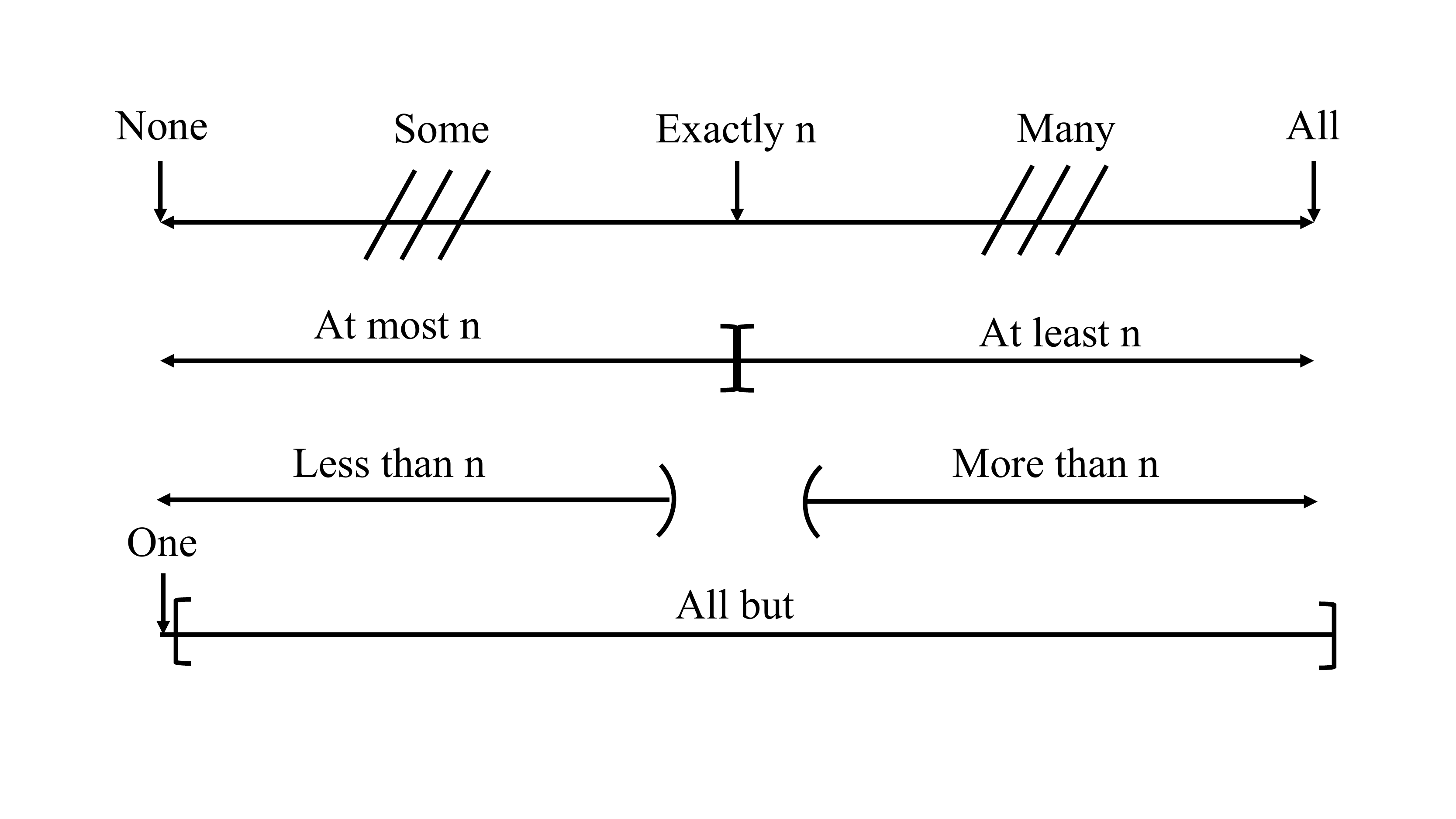}
	\caption{Quantifier Scopes}
	\label{fig:quantifierscopes}
\end{figure}

Determiners are frequent parts of speech in the English language. While determiners in general describe what a noun refers to, for instance, ``the'', ``some'', or ``their'', quantifiers represent a subcategory of determiners referring to a certain quantity of the noun.
Keenan and Stavi offer an extensive list of natural language determiners, which includes a substantial number of quantifiers~\cite{Keenan1986-KEEASC}. 
Many quantifiers have similar meaning. As an example, ``at least n'' or ``n or more'' include the same set of items with regard to ``n''. 
We categorized the quantifiers according to their semantic scope, i.e.  quantifiers of the same category hold true for equal sets. Based on this categorization, we defined 11 scopes, of which two defined as \texttt{Some} and \texttt{Many} are ambiguous and thus irrelevant to this paper, which deals with explicit quantifiers. 
The 9 exact scopes are: \texttt{None}, \texttt{All}, \texttt{Exactly n}, \texttt{At least}, \texttt{At most}, \texttt{Less than}, \texttt{More than}, \texttt{One} and \texttt{All but}, as depicted in \figurename~\ref{fig:quantifierscopes}.
Exact quantifiers are either numbers, like ``one'', ``two'', or ``exactly a hundred'', which is contained in the scope \texttt{Exactly n}. When speaking of \texttt{All}, every element of a set is included, while \texttt{None} as its counterpart excludes all elements. Some quantifiers are graded: The scope \texttt{At least} is upward entailing, i.e. it includes all elements in the subset $[n, max)$, while the scope \texttt{At most} is its downward entailing counterpart. The scopes \texttt{More than} and \texttt{Less than} are similar, however, they have open intervals, thus exclude the value ``n''. 
The scope \texttt{One} refers to a certain instance, rather than any set with a certain property. Quantifiers included in this scope are, for example, ``the'', or ``a'', as in ``the object'', or ``a group of objects''. The scope \texttt{All but} is the counterpart to this scope, excluding this instance of a set.

The quantifiers listed by Keenan and Stavi~\cite{Keenan1986-KEEASC} can be classified into these semantic scope categories. From the original set of determiners, indefinite quantifiers, such as ``nearly all'' are excluded and duplicates or similar quantifiers, such ``five or more'' and ``a hundred or more'' are aggregated into one scope. 

Hence, natural language possesses a variety of determiners~\cite{Glanzberg2006} that can be utilized to express the different scopes of quantification. This presents us the question, whether some determiners are more readable and comprehensible than others with an equivalent semantic scope. 

\subsection{Affirmative and Negative Sentences}
\label{sec:2:Negative Sentences}

Christensen~\cite{Christensen2009} examined the neurobiological implications of affirmative and negative sentences in the brain. The findings suggest that different brain areas are activated when processing affirmative and negative sentences, i.e. sentences containing a negative operator. Moreover, the brain requires more processing time for negation than for affirmation, thus response time is also longer. Performance, however, is suggested to be equal for both types of sentences. According to Christensen, affirmative sentences involve a simpler semantic and syntactic structure than negative sentences~\cite{Christensen2009}.
According to their work, negative sentences entail a more complex syntactical structure, which requires ``additional syntactic computation'' in the brain~\cite{Christensen2009}. Christensen denotes affirmative polarity as ``default'', to which negative operators add additional structure. More precisely, when reading negative sentences, the human brain interprets all sentences as affirmative at first and in the second step, adds negative polarity to negative sentences~\cite{Christensen2009}.

In the dataset of requirements specifications used as a source for this paper, quantifiers are formulated in both, affirmative and negative form. To express the same semantic scope, one can employ positive and negative structures. For example, one could say ``at least ten'', or equivalently ``no fewer than ten''. Which of these two possibilities is more advisable to use in requirements specifications? Although Christensen has given an indication on the answer to this question, it could also be assumable that negative quantifiers yield longer response time, but better reading comprehension. 

\subsection{Requirements Readability}
\label{sec:3:Research Questions}
Requirements artifact quality can be understood as the extent to which properties of the artifact impact activities that are executed based on the artifact. In particular, quality factors affect the effectiveness and efficiency of use~\cite{Femmer19}. One relevant activity on requirements specifications is reading~\cite{ATOUM2018}. Consequently, good quality in practice includes efficient and effective readability of the requirements specifications. We therefore examine the implications of the quality factor quantifiers on effectiveness and efficiency of reading. 
We understand readability as an indicator for the ``ease of understanding or comprehension due to the style of writing''~\cite{Klare:2000}. Readability thus describes reading efficiency and good quality in readability minimizes the reading effort to gain comprehension of the requirements.
Reading comprehension indicates effectiveness of reading. When considering readability, the reading performance must not be neglected. Although ease of understanding and sentence comprehension are closely related, good readability that yields a wrong understanding of the phrase is an indicator of bad quality. It is thus required to achieve both, efficiency and effectiveness in requirements specifications.

Objective indicators are one aspect of the assessment of quality in readability. Klare \cite{Klare:2000} makes a point with the statement: ``The reader must be the judge''. Hence, subjective perception should also be considered in regard to readability of requirements specifications.
Therefore, we examine the readability, comprehension, and subjective perception of syntactically affirmative and negative quantifiers for a limited set of quantifier scopes.

\section{Study Design}
In this study, we analyze the impact of affirmative and negative quantifier phrases on readability, comprehension, and perceived difficulty.

\textbf{Research Question 1:}
How does affirmative and negative syntax of quantifiers impact reading efficiency?

\textbf{Research Question 2:} 
How does affirmative and negative syntax of quantifiers impact reading effectiveness?

\textbf{Research Question 3:} 
How does affirmative and negative syntax of quantifiers impact the subjective perception of reading difficulty?








\subsection{Data Collection}
\label{chp:Experiment}
We conducted an experiment to gather data on the research questions following the guideline by Wohlin~et~al.~\cite{Wohlin2000}. 
To assess the differences between affirmative and negative quantifiers, we examine the relationship between quantifier syntax and readability, comprehension, and perceived difficulty. 
%
%
We implemented a web-based experiment, which yields a controllable testing environment and allows for a general evaluation of our hypotheses since the experiment questions are not bound to a certain context and thus do not require prior knowledge on a particular topic. Instead, the web application contains an artificial problem to easily gain first results on the research questions.




\subsection{Study Objects and Treatments}
Based on the research questions, the independent variable that is controlled in this experiment is the syntactical structure of the quantifying sentences. The two treatments are \emph{affirmative} and \emph{negative} syntactical structure. The dependent variables that will be measured in the experiment are the readability, understandability, and subjective perception of difficulty for each treatment.

Wohlin et al.~\cite{Wohlin2000} offer a standard design type for such experiments with one factor and two treatments. Leaning on this design type, we aim ``to compare the two treatments against each other''. 
Furthermore, we choose a paired comparison study design, where ``each subject uses both treatments on the same object''~\cite{Wohlin2000}.
We compare the two treatments, affirmative and negative syntactical sentence structure, on sentences addressing the same quantifier scope. 

Table~\ref{tab:quantifiersamples} lists all samples of quantifiers that are given in the experiment. These samples were made up by us and did not have any specific background or focus. For each quantifier scope, an affirmative quantifier and a negative equivalent is displayed. Note that the scope \texttt{None} is a special case, as it is naturally negative and thus its counterpart is positive. Words in bold are characteristic for the respective syntactic structure. 

\begin{table}
\centering
\caption{Affirmative and negative syntax samples for each quantifier scope.}
\label{tab:quantifiersamples}
\begin{tabularx}{\textwidth}{@{}l@{\hskip 1em}X@{\hskip 1em}X@{}}
\toprule
\textbf{Scope} & \textbf{Affirmative Syntax Sample} & \textbf{Negative Syntax Sample}\\
\midrule
	All & \textbf{All} registered machines must be provided in the database. 
	    & \textbf{No} deficit of a machine is \textbf{not} provided in the database \\ [1ex]
	None & \textbf{All} access is blocked \textbf{without} a valid login.  
	    & \textbf{None} of the service workers may have access to 'Budget'. \\[1ex]
	More than & At \textbf{more than} 5 deficits the signal token turns red. 
	    & \textbf{Not only} defective machines are displayed in the system. \\[1ex]
	At least & The number of new parts per order must be \textbf{at least} 3. 
	    &  A highly defective machine has \textbf{no less than} 5 defects. \\[1ex]
	At most & Per machine, \textbf{at most} 4 photos can be uploaded to the database. 
	    & An approved machine has \textbf{no more than} 2 minor defects. \\ [1ex]
	Less than & \textbf{Less than} 3 supervisors may be assigned to each service worker. 
	    & \textbf{Not as many} supervisors \textbf{as} 3 may be assigned to each machine. \\ [1ex]
	Exactly n & \textbf{Exactly} 2 emergency contacts must be displayed at all times. 
	    & \textbf{No more or less than} 2 supervisors must be online at all times.\\[1ex]
	All but & \textbf{All} machines \textbf{but} the current one must be on the list 'new jobs'. 
	    & \textbf{No} location \textbf{but} the location of the current machine is on the map. \\[1ex]
	One & \textbf{Only} the location of the current machine is on the map. 
	    & The current job is the \textbf{only} job that is \textbf{not} listed in 'last jobs' \\[1ex]
\bottomrule
\end{tabularx}
\end{table}

\begin{figure}
    \centering
    \subfloat[Image 1 (correct answer)]{{\includegraphics[width=4cm]{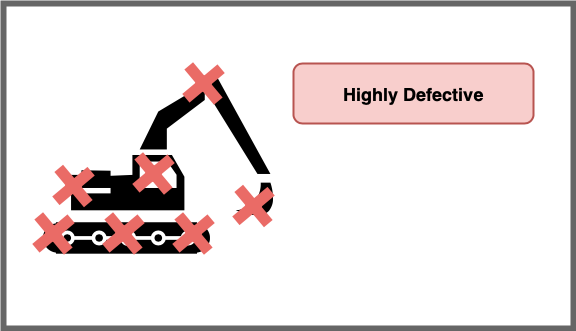} }}%
    \qquad
    \subfloat[Image 2 (correct answer)]{{\includegraphics[width=4cm]{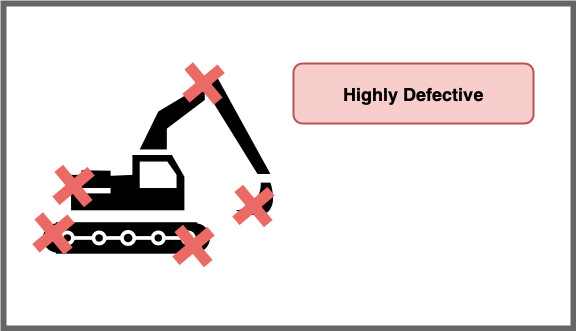} }}%
     \qquad
    \subfloat[Image 3 (wrong answer)]{{\includegraphics[width=4cm]{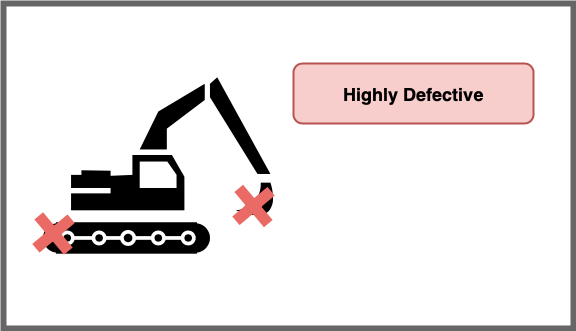} }}%
    \caption{Example question from the experiment: Which of the images match the sentence ``A highly defective machine has no less than 5 defects''?}
    \label{fig:exampleImages}%
\end{figure}

The task of the experiment was to compare the given sentence with three given situations and decide which of the three situations (one, two, or all three) match the given sentence. The situations are presented as images. \figurename~\ref{fig:exampleImages} depicts one of the 18 answers in the experiment and belongs to the sentence ``A highly defective machine has no less than 5 defects''. The images are nearly identical, except for quantification, represented in red crosses in this image. The quantifier scope \texttt{At least}, which is stated here in negative syntax, entails the amounts of \{five, six, seven, \dots \} crosses. Thus, the correct answers to select are Image 1 and Image 2, as they are entailed, whereas Image 3 does not accurately describe the sentence. 

As recommended by Wohlin et al.~\cite{Wohlin2000}, the order of the sentences is randomized to prevent the effect of order and have a balanced design, such that the subjects' paths through the experiment are diverse. To further avert information gain from past questions, we not only randomized each sentence pair, but mix all sentences. Moreover, we created sentence pairs with identical quantifying scopes and similar but not equal semantic meaning (see Table~\ref{tab:quantifiersamples}). 





\subsection{Subject Selection}
We selected the subjects for the experiment by convenience sampling~\cite{Wohlin2000} via mailing lists, or personal and second-degree contacts of the authors. The experiment was conducted online with anonymous participants. Thus, we had no control over the situation and context in which the experiment was executed by each participant.
Our web-based experiment was started by 76 participants of which 51 completed the experiment. All figures in this paper refer to the 51 participants that completed the experiment. Prior to the experiment, we ask the participants whether they have a background in computer science (\textit{yes}: 94.1\%, \textit{no}: 5.9\%), whether they are native English speakers (\textit{yes}: 5.9\%, \textit{no}: 94.1\%), and what their profession is (\textit{academic}: 23.5\%, \textit{professional}: 49.0\%, \textit{student}: 27.5\%).

\subsection{Data Analysis}
\label{sec:4:Metrics}

To answer the proposed research questions, we selected the following metrics. 

\textbf{Readability:} To evaluate readability in terms of efficient reading, the effort of reading needs to be measured. 
Many studies and experiments measure \emph{reading time} as an indicator for the level of difficulty it requires to process a sentence~\cite{MacKay1966,CIRILO1980,GRAESSER1980}. Therefore, we also use reading time as an indicator of reading difficulty to examine the effort for a person to understand a sentence.
In the experiment, we measured the time that a participant required to read and comprehend the sentence. The counter was started when the sentence appeared on the screen and stopped again when the participant clicked on the button to submit the answer. 
To examine the differences in reading time between the affirmative and the negative syntax sample for each scope, we applied a Wilcoxon signed-rank test, which is suitable for comparing two paired samples with data that is not normally distributed. As we will see later, the assumption of the Wilcoxon signed-rank test holds, since the reading times in our experiment are not normally distributed. The test returns a \emph{p-value} to assess the significance of the effect and an \emph{effects size} to assess the magnitude of the effect. 

\textbf{Understandability:} 
To measure correctness, we test whether the understanding reflects the true meaning of the sentence or represents a false belief. As discussed in Section~\ref{sec:1:Quantifiers}, some quantifiers entail a range of correct solutions. For instance, the quantifier \texttt{five items or more} entails all numbers of items of five and above (i.e. five, six, seven,\dots). For other quantifiers, like \texttt{exactly five items}, one number, namely five, is the correct quantification, while all other numbers, like four or six items, do not reflect the true meaning of the quantifier. 
Hence, the three situations presented as answers in the online experiment are independent and include correct as well is incorrect quantifications of the given statement.
We consider a sentence as ``understood'' if all included and excluded options are correctly identified.
To examine the differences in correctly understood sentences, we build a 2x2 contingency table containing the number of participants with correct and incorrect answers in affirmative and negative sentences (see Table~\ref{tbl:cTable}). Since our samples are matched, we focus on the discordant cells in the contingency table ($b$ and $c$) and apply an exact binomial test to compare the discordant cell $b$ to a binomial distribution with size parameter $n = b + c$ and probability $p = 0.5$. This test is suggested for 2x2 contingency tables with matched samples and few samples in the discordant cells ($b+c<25$). As a measure for the effect size, we report the \emph{odds ratio}: $\mathit{OR}=b/c$.

\begin{table}
    \centering
    \caption{2x2 contingency table of correct and incorrect answers for one scope.}
    \label{tbl:cTable}
\begin{tabular}{ rr|c|c| }
\multicolumn{1}{r}{}
&\multicolumn{1}{r}{}
 & \multicolumn{2}{c}{\textbf{affirmative syntax}} \\
\multicolumn{1}{r}{}
&\multicolumn{1}{r}{}
 &  \multicolumn{1}{c}{incorrect}
 & \multicolumn{1}{c}{correct} \\
\cline{3-4}
\textbf{negative syntax} & incorrect & a & b \\
\cline{3-4}
& correct & c & d \\
\cline{3-4}
\end{tabular}
\end{table}

\textbf{Perceived Difficulty:} 
For the determination of perceived difficulty, we asked the participants to rate the reading difficulty on a scale with the values ``easy'', ``medium'', and ``difficult''. We use this ordinal scale as it allows for the assessment of less to greater, where intervals are not equal. 
The perceived difficulty is subjective and intervals between the options ``easy'' and ``medium'', as well as between ``medium'' and ``difficult'' are not necessarily equal. Furthermore, levels of difficulty may differ in between the category itself.
To examine the differences in the perceived difficulty, we applied a Wilcoxon signed-rank test, which is suitable for comparing two paired samples with ordinal data.


\subsection{Experiment Validity}
\label{sec:3:ValidityThreats-Experiment}
Prior to starting the experiment and collecting the data, we launched a test run with three participants to receive feedback on the correctness of language, the comprehensibility of the overall experiment, and remaining technical bugs.
Although the affirmative and negative syntax sample for each quantifier describe different situations (see Section~\ref{chp:Experiment}), the generated sentences are similar by choosing a narrow vocabulary throughout the experiment. The difference between sentences averages about 1.77 words, where in five cases the affirmative sentence contains more words and in four cases the negative sentences is longer. The sentences have a simple structure, such that other syntactical phenomena, like sentence complexity, do not invalidate the results. On average, the sentences have 11 words.
For each sentence, the study subjects have three answer options. To avoid complexity of the answers through e.g. answer sentences that are difficult to understand, the answer options are displayed as images (see \figurename~\ref{fig:exampleImages}). Like the sentences, the images have a similar image vocabulary containing equal symbols and language of form. For each sentence in the experiment, the images have minimal, but distinguishable differences. One or more of these images represent the correct meaning of the sentence given. By providing more than one correct answer, the effect of exclusion by comparison between different images should be avoided and the subject is forced to deal with each answer option separately. 

To assure transparency and improve reproducibility, we have published the raw results of the experiment and the R-script that we used for processing the data.\footnote{\url{https://doi.org/10.6084/m9.figshare.10248311}}





\section{Study Results}
\label{chp:Results}

\subsection{Effects on Readability (RQ1)}
\label{sec:1:Readability}

\begin{figure}
	\centering
	\includegraphics[width=\textwidth]{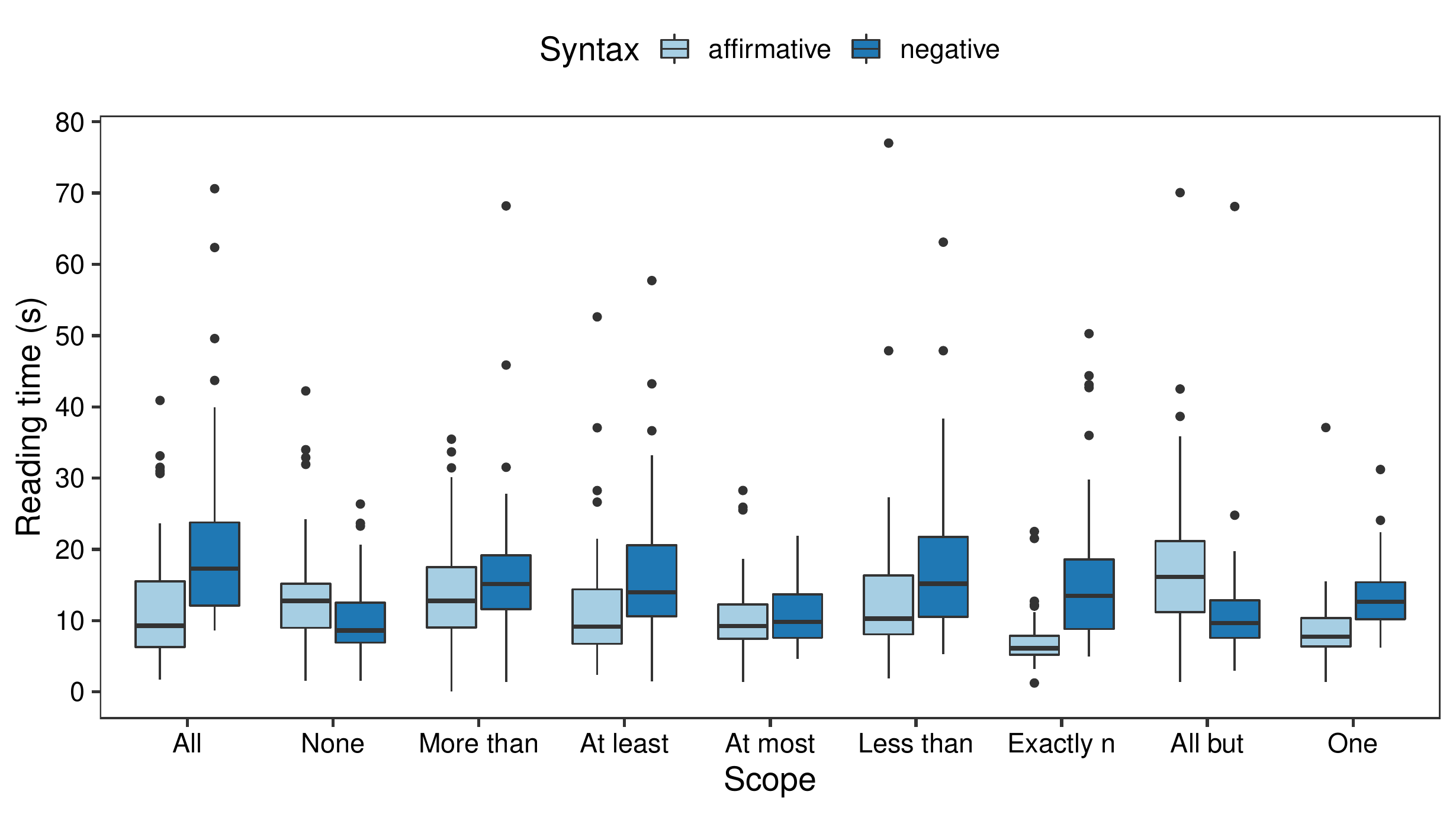}
	\caption{Distribution of reading times per scope}
	\label{fig:timeresult}
\end{figure}

When examining the collected reading times, we saw that all values were below 77 seconds, except for two data points where the reading time were 665 seconds and 12,281 seconds (both measured for sentences with negative syntax). Since we had no control over the situation in which the experiment was conducted, we consider both data points as outliers, possibly due to a disturbance of the participant, and removed the data points as well as their corresponding affirmative sentences from the dataset. 
\figurename~\ref{fig:timeresult} displays boxplots of the remaining reading times for each scope. 
As shown in the figure, for six of the nine pairs, it took more time on average to read the negative quantifier compared with the positive quantifier of the same scope. 

Table~\ref{tab:time-test} lists the results of the Wilcoxon signed-rank test for each scope in terms of the p-value and effect size for significance level $\alpha=0.05$.

\begin{table}
\centering
\caption{Wilcoxon signed-rank test for differences in reading times between affirmative and negative syntax in each scope.}%
\label{tab:time-test}
\begin{tabularx}{\textwidth}{lRRRRRR@{\hskip 1em}R@{\hskip 1em}RR}
	\toprule	
		\textbf{Scope} & All & None & More than & At least & At most & Less than & Exactly n & All but & One \\
	\midrule 
		\textbf{p-value} & \textbf{.000} & \textbf{.019} & .077 & \textbf{.000} & .409 & \textbf{.001} & \textbf{.000} & \textbf{.000} & \textbf{.000}  \\
		\textbf{effect size} & .46 & .23 & .18 & .37 & .08 & .32 & .58 & .52 & .52  \\
	\bottomrule
\end{tabularx}
\end{table}

According to the significance test, the following quantifier scopes exhibit a significant difference in reading time: \texttt{All}, \texttt{None}, \texttt{At least}, \texttt{Less than}, \texttt{Exactly n}, \texttt{All but}, and  \texttt{One}.
Only for quantifiers \texttt{At most}, and \texttt{More than}, we were not able to reject the null hypothesis of equal reading times.
Among the quantifier scopes, \texttt{All but} and \texttt{None} yield significantly longer reading times for the affirmative quantifier than for the negative, as depicted in \figurename~\ref{fig:timeresult}. In all other cases, affirmative quantifiers perform better than their negative equivalences regarding the average reading time. The effect size values indicate small ($0.2$) to moderate effects ($0.5$)~\cite{Cohen2013}. An effect size of $0$ means that exactly 50\% of participants spent less reading time for the affirmative sentence than the mean reading time for the negative case (i.e., there is no difference). A moderate effect size of $0.5$ indicates that 69\% of participants spent less reading time for the affirmative sentence than the mean reading time for the negative case, while for large effect size ($0.8$) this is already true for 79\% of participants.

\subsection{Effects on Comprehension (RQ2)}
\label{sec:2:Comprehension}

\begin{figure}
    \centering
    \includegraphics[width=\textwidth]{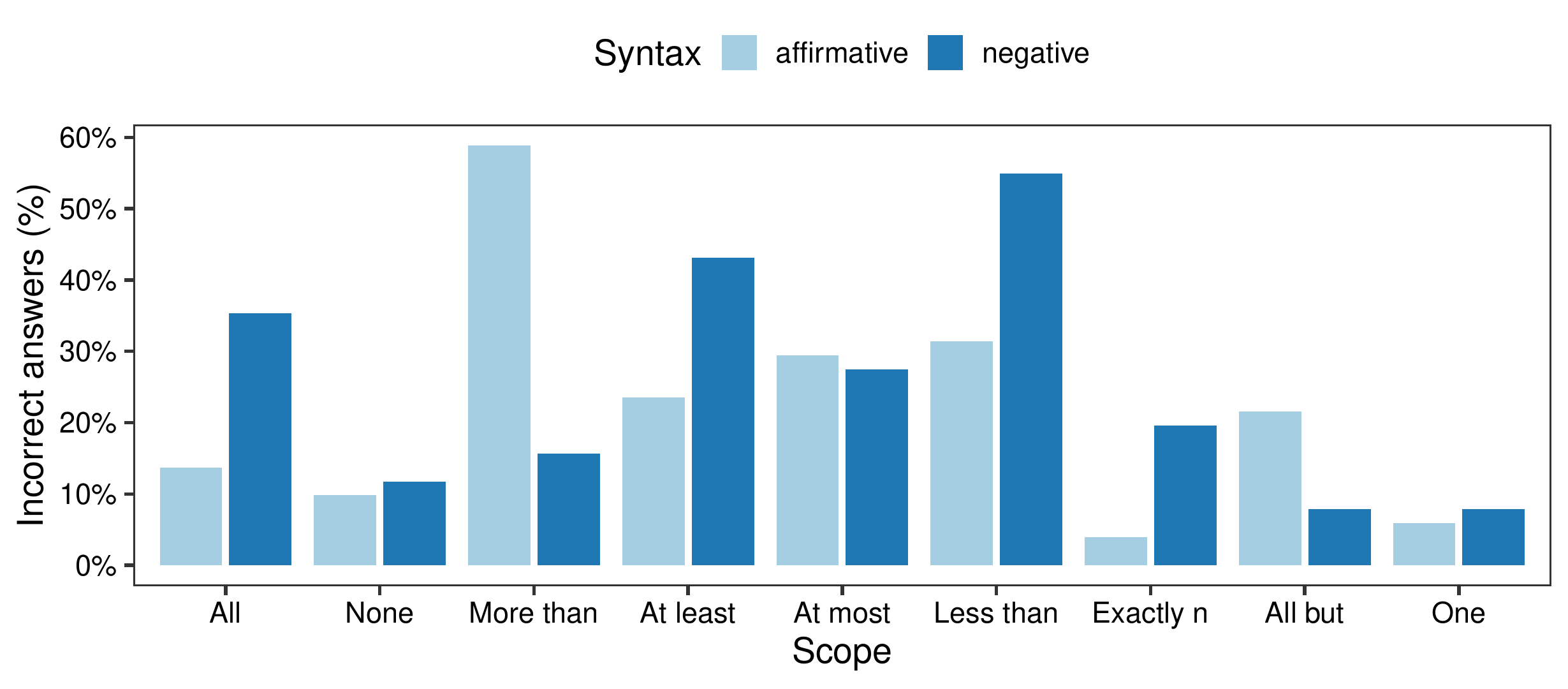}
    \caption{Distribution of incorrect answers per scope}
    \label{fig:erroresult}
\end{figure}

\figurename~\ref{fig:erroresult} shows the ratio of incorrect answers per scope. For 6 of the 9 quantifiers, our participants made more errors in the sentence with negative syntax. Only for the quantifier scopes \texttt{More than}, \texttt{At most}, and \texttt{All but}, the participants made more errors in the sentence with affirmative syntax.

Table~\ref{tab:error-test} lists the results of the exact binomial test for each scope in terms of the p-value and the odds ratio as a measure for effect size for significance level $\alpha=0.05$.

\begin{table}
\centering
\caption{Binomial test for differences in error ratio between affirmative and negative syntax in each scope}%
\label{tab:error-test}
\begin{tabularx}{\textwidth}{lRRRRRR@{\hskip 1em}R@{\hskip 1em}RR}
	\toprule	
		\textbf{Scope} & All & None & More than & At least & At most & Less than & Exactly n & All but & One \\
	\midrule 
		\textbf{p-value} & \textbf{.007} & 1.0 & \textbf{.000} & \textbf{.013} & 1.0 & \textbf{.017} & \textbf{.008}& \textbf{.016} & 1.0   \\
		\textbf{odds ratio} & 6.50 & 1.50 & 12.50 & 6.00 & 1.33 & 3.40 & Inf & Inf & 2.00  \\
	\bottomrule
\end{tabularx}
\end{table}

For 5 of the 9 quantifier scopes, our participants made significantly more errors in the negative sentence. For the scopes \texttt{All but} and \texttt{More than}, our participants made significantly more errors in the affirmative sentences. The odds ratios as a measure for effect size varied between small effects ($\mathit{or}<2.57$), moderate effects ($2.75\leq\mathit{or}<5.09$) and large effects ($\mathit{or}\geq5.09$)~\cite{Chen2010}. An odds ration of 6.5 for the scope \texttt{All}, for example, means that the chances of incorrectly answering the negative sentence was 6.5 times higher than the chances of answering the affirmative sentence incorrectly. 

\subsection{Effects on Perceived Difficulty (RQ3)}
\label{sec:3:Self-Assesment}

\begin{figure}
	\centering
	\includegraphics[width=\textwidth]{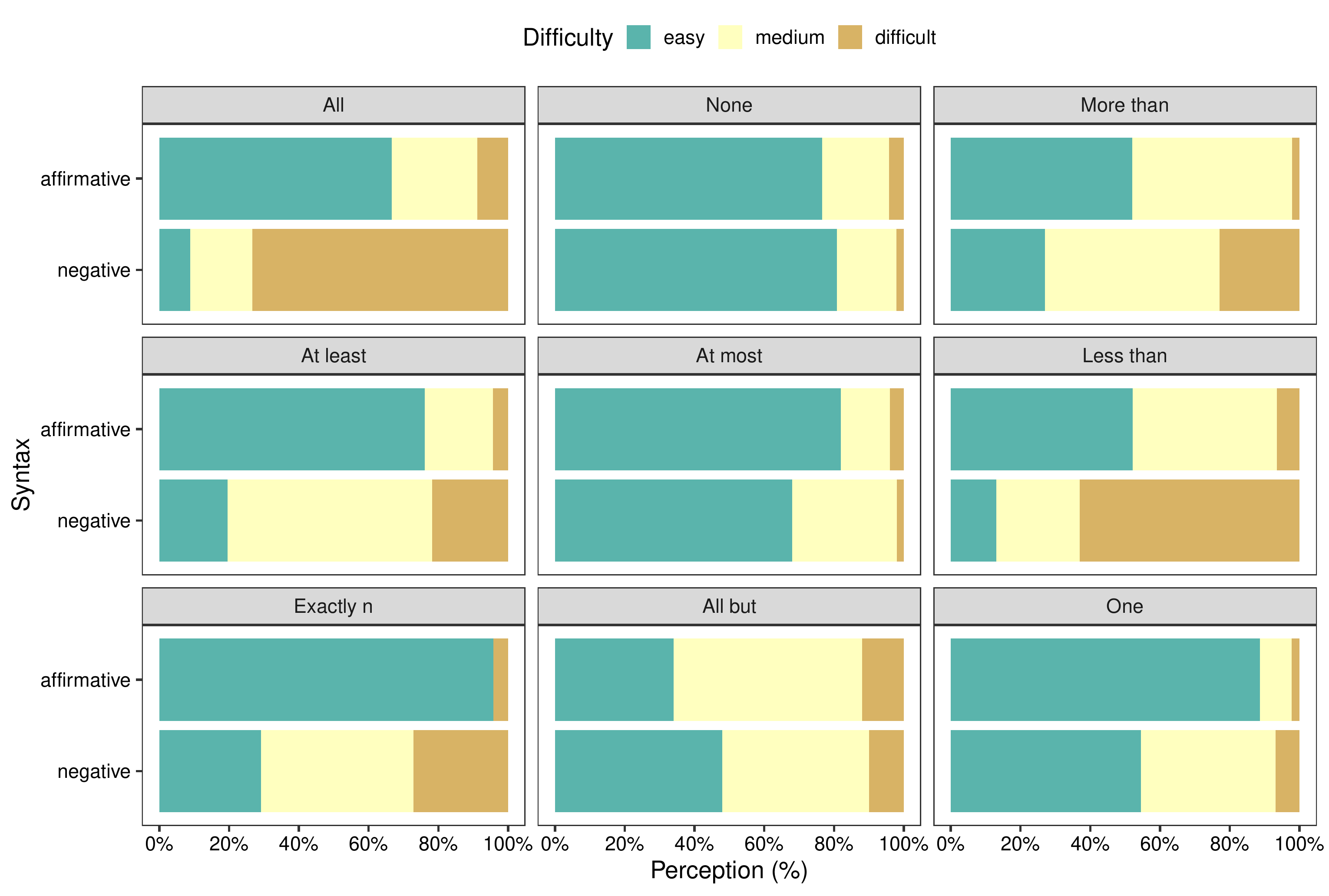}
	\caption{Subjective perception of sentence difficulty per scope}
	\label{fig:selfassessment-result}
\end{figure}

After each question in the experiment, the participants were confronted with a self-assessment scale on how difficult they perceived the sentences. Answer options were \texttt{easy}, \texttt{medium}, and \texttt{difficult}. \figurename~\ref{fig:selfassessment-result} depicts the assessments over the participants. 

Table~\ref{tab:perception-test} lists the results of the Wilcoxon signed-rank test for each scope in terms of the p-value and effect size for significance level $\alpha=0.05$.

\begin{table}
\centering
\caption{Wilcoxon signed-rank test for differences in perceived difficulty between affirmative and negative syntax in each scope}%
\label{tab:perception-test}
\begin{tabularx}{\textwidth}{lRRRRRR@{\hskip 1em}R@{\hskip 1em}RR}
	\toprule	
		\textbf{Scope} & All & None & More than & At least & At most & Less than & Exactly n & All but & One \\
	\midrule 
		\textbf{p-value} & \textbf{.000} & .510 & \textbf{.000} & \textbf{.000} & .141 & \textbf{.000} & \textbf{.000} & .164 & \textbf{.001}  \\
		\textbf{effect size} & .55 & .07 & .36 & .51 & .15 & .49 & .52 & .14 & .37  \\
	\bottomrule
\end{tabularx}
\end{table}

Six of the nine quantifier scopes show significant differences in the perceived difficulty. For all of these scopes, the participants perceived the affirmative phrase as easier. The effect size measures for the scopes with significant differences all indicate moderate effects ($0.35\leq \mathit{effect} < 0.65$)~\cite{Cohen2013}.
For the remaining three scopes, the difference in perceived difficulty is not significant.

\subsection{Summary of the Results}
\label{sec:4:Summary}

\begin{figure}
    \centering
    \includegraphics[width=\textwidth]{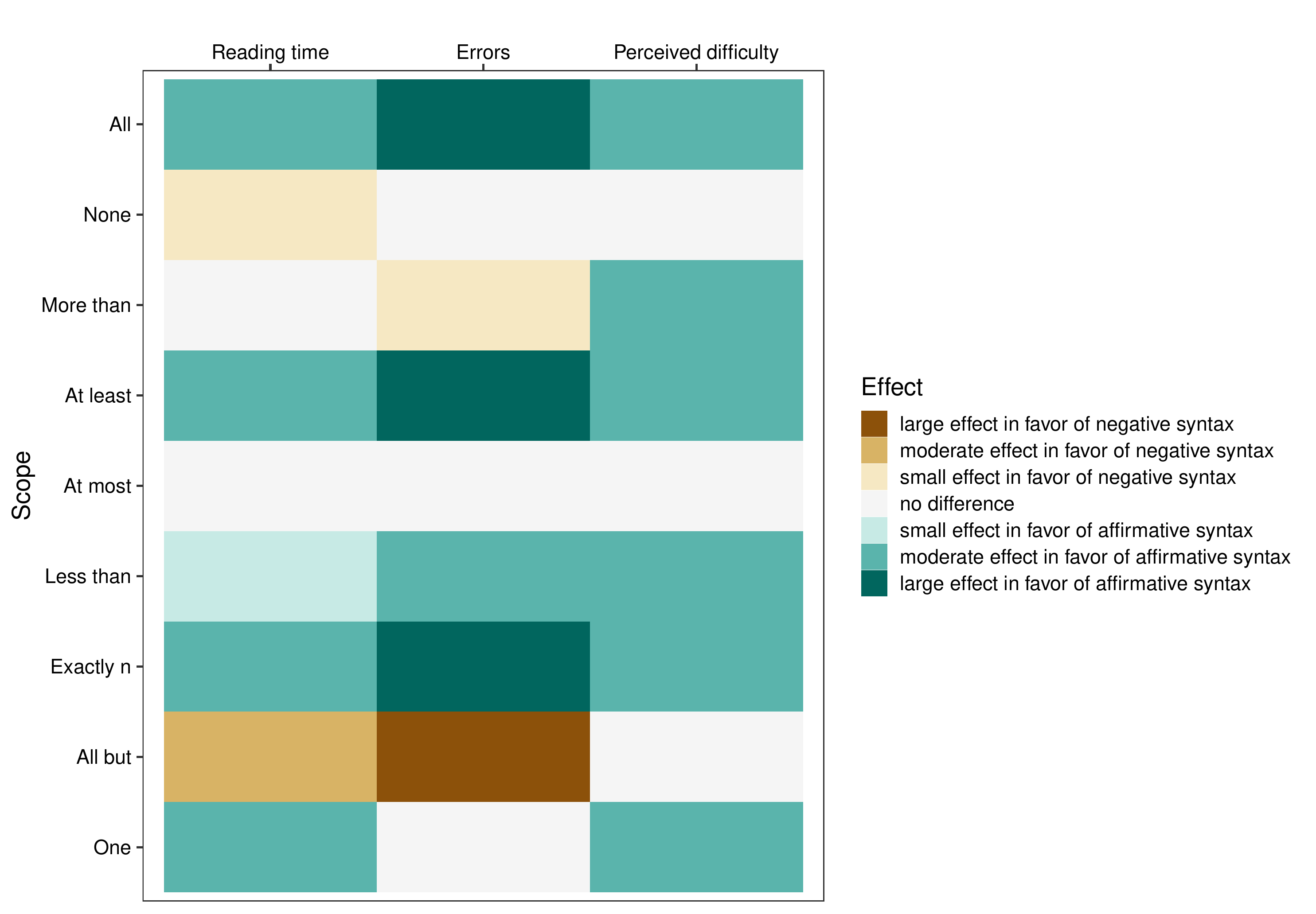}
    \caption{Summary of results}
    \label{fig:summary}
\end{figure}

\figurename~\ref{fig:summary} summarizes the results of the three research questions. The figure shows the scopes and the measured differences with a qualitative evaluation of the effect sizes according to Cohen~\cite{Cohen2013}.

Overall, negative quantifiers perform worse in more cases than positive quantifiers, which is clear for the quantifiers \texttt{All}, \texttt{At least} and \texttt{Exactly n} and also apparent for the quantifier \texttt{Less than}.
For other quantifiers, results are neutral, like the quantifier \texttt{None}, which is in a special position, as it naturally is formulated in negative syntax, or \texttt{At most} and \texttt{One}, which exhibit neutral objective measurements, but show tendencies in self-assessment towards differences in subjective difficulty.
The quantifier \texttt{More than} was the outlier in the measurements regarding the number of mistakes in the positive sentence. Thus, this result should be treated with care. Especially, since self-assessment showed clear tendencies that the negative sentence is more difficult.
Last but not least, the extra quantifier \texttt{All but} performed worse in two measurements, namely reading time and self-assessment, and only neutral when it came to the number of wrong answers. Hence, it is the only quantifier that yields a worse overall performance of the positive quantifier. 

\section{Discussion}
\label{chp:Conclusion}

\subsection{Threats to Validity}
\label{sec:1:Validity}
For the interpretation of the results, several threats to validity need to be considered.

\textbf{Construct Validity:}
Only one specific representative quantifier was stated for each scope in the experiment. Thus, results are inferred from these exact representatives, not the quantifier scopes in general. Using other quantifiers to express a scope may possibly yield different results. 
In addition, results may depend on the setup of the experiment. 
As subject area of all sample requirements, we used a software product for machine maintenance.
Each quantifier was then embedded in a sentence that was equal for all test subjects and had specific answer options encoded as images. A different use of sentences, images, or other factors, such as the professional background and English proficiency of the subjects, could lead to different results. 

\textbf{Conclusion and Internal Validity:} 
Prior to starting the experiment and collecting the data, we launched a test run with three participants to receive feedback on the correctness of language, the comprehensibility of the overall experiment, and remaining technical bugs.
Since we used an experimental design where all participants were faced with all treatments in a random order, we do not expect effects due to a confounding variable. The sample size of 51 subjects is reasonable to draw statistical conclusions. We only asked the participants whether they have a background in computer science, whether they are native English speakers, and what their profession is. We did not analyze the effect of these demographic factors due to the small size of single groups. In addition, we are not able to analyze effects related to further contextual factors of subjects such as experience, closeness to the application domain, or others. 
We selected the applied statistical tests based on the characteristics of the experiment (e.g., paired samples) and checked the test's assumptions (e.g., normal distribution). All elicited measures (reading times, number of errors, and perceived difficulty) are independent from any kind of judgement by the authors.
To make the results transparent, we report p-value and effect size. Still, we used an arbitrary, yet common, significance level threshold of $\alpha=0.05$ for the statistical tests.

\textbf{External Validity:}
Since we used convenience sampling, we cannot claim that our participant group is representative for the group of all people working with requirements. Particularly, participants with a different language background may have more or less difficulties with negative or affirmative syntax. 
In addition, we used artificial requirements for the treatments. We cannot claim that these are representative for real requirements in the context of each participant.

\subsection{Interpretation and Writing Guidelines}
\label{sec:2:Interpretation}

Taking the threats to validity into account, we can cautiously interpret these results. We conclude that some quantifiers exhibit better readability and better comprehension when phrased in affirmative syntax. 
Furthermore, self-perception mostly coincides with readability and comprehension, which might be owed to the fact that longer reading time and the guessing of answers impact the perceived difficulty of the sentences. Nevertheless, even for the quantifier scopes where participants made significantly more errors and spent more reading time with the affirmative syntax, the participants did not perceive the negative phrasing as easier to read (see \figurename~\ref{fig:summary}).

An observation that was surprising to us was the high error rate for the scopes \texttt{More than} (affirmative case) and \texttt{Less than} (negative case). As shown in \figurename~\ref{fig:erroresult}, almost 60\% of our participants answered the question incorrectly. A deeper analysis of the results showed that, for the sentence ``more than x\ldots'', a large number of participants incorrectly selected the answer that showed exactly x instances. This results is mirrored in the negative case of the \texttt{Less than} scope. Apparently, our participants had difficulties with sentences that represent open intervals. Given that the error ratio for scopes \texttt{At least}, and \texttt{At most} is lower, we may conclude that it is better to use formulations that represent closed intervals.

In summary, we draw the following conclusions that can be used as advice for writing requirements that are faster to read, have lower chances of misinterpretation, and are perceived as easier to read:

\fbox{
\begin{minipage}{0.92\textwidth}
\begin{compactenum}
    \item Use \textbf{affirmative} syntax for scopes \texttt{All}, \texttt{At least}, \texttt{Less than}, \texttt{Exactly n}, and \texttt{One}:
    \begin{compactitem}
        \item Write \texttt{All\ldots} instead of \texttt{No\ldots not}
        \item Write \texttt{At least\ldots} instead of \texttt{No less than\ldots}
        \item Write \texttt{Less than\ldots} instead of \texttt{Not as many as\ldots}
        \item Write \texttt{Exactly n\ldots} instead of \texttt{No more or less than n\ldots}
        \item Write \texttt{Only \ldots} instead of \texttt{Only\ldots not}
    \end{compactitem}
    \item Use \textbf{negative} syntax for the scope \texttt{None}:
    \begin{compactitem}
        \item Write \texttt{None of\ldots} instead of \texttt{All\ldots without}
    \end{compactitem}
    \item Use \textbf{closed-interval} formulation instead of open-interval formulation:
    \begin{compactitem}
        \item Write \texttt{At least\ldots} instead of \texttt{More than\ldots}
        \item Write \texttt{At most\ldots} instead of \texttt{Less than\ldots}
    \end{compactitem}
    \item In doubt, use \textbf{affirmative} syntax since it is perceived as easier.
\end{compactenum}
\end{minipage}
}

\subsection{Relation to Existing Evidence}
Berry and Kamsties~\cite{Berry05} noticed that some quantifiers may be dangerous to use in requirements because they create ambiguity. They specifically recommend avoiding indefinite quantifiers, such as ``nearly all'', and the quantifier \emph{all} with a plural noun because it is not clear whether the corresponding statement applies to each instance separately or to all instances as a whole. In our experiment, the affirmative sentence for the scope \texttt{All} contained the quantifier \emph{all} with a plural noun. Although 10\% of our participants gave an incorrect answer for this sentence, this number was not particularly higher than in other scopes. 

Christensen~\cite{Christensen2009} performed an empirical study on the effect of negative and affirmative statements on response time (i.e., how fast did subjects answer questions about the presented statements) and reading performance (i.e., how often were the answers correct). They found significantly shorter response times for affirmative sentences and lower error rates (differences were not significant). Our results corroborate the results of Christensen in general although there were some scopes with effects in favor of the negative syntax (e.g., \texttt{All but}).

\section{Conclusion}
\label{sec:3:Conclusion}

In the course of this study, we raised questions on the readability, comprehension, and subjective difficulty of affirmative and negative quantifier formulations in natural language requirements. We designed and conducted a web-based experiment, from which we evaluated the results using the time for readability, correctness for comprehension, and self-assessment for subjective difficulty. The results were interpreted and yielded a tendency towards better overall performance of affirmative quantifiers compared to their negative equivalences. 
This extends and confirms related studies from psycholinguistics. Moreover, our results suggest using quantifiers representing closed intervals instead of open intervals.

Our results depict first empirical impressions on quantifiers in requirements specifications. However, much about this topic remains to examine.
First of all, it remains to review, whether the categorization of quantifiers in this study is sensible or whether other categorizations are also possible. 
Since we only examined one concrete quantifier formulation for each scope, the results may not be generalized to other syntactic representations of the same scope. Future research could thus involve repeating the experiment with a different set of quantifiers in a different context to validate the results and give additional information to eventually generalize the results. 
Last but not least, certain quantifiers could be proposed as new requirements smells and tools may be used to detect these smells to improve the quality of natural language requirements.

\bibliographystyle{splncs04}
\bibliography{references}

\end{document}